\documentclass[12pt,preprint]{aastex}
\usepackage{flushrt,float}

\slugcomment{Accepted for publication on The Astrophysical Journal}
\shortauthors{Chieffi, Limongi} \shorttitle{Massive Stars}

\begin{document}
 
\title{Explosive yields of massive stars from $\rm \bf Z=0$ to $\rm \bf Z=Z_\odot$}

\author{Alessandro Chieffi\altaffilmark{1,3,4} and Marco Limongi\altaffilmark{2,3,4}}

\altaffiltext{1}{Istituto di Astrofisica Spaziale e Fisica Cosmica (CNR), Via Fosso del Cavaliere, I-00133, Roma, Italy;
achieffi@rm.iasf.cnr.it}

\altaffiltext{2}{Istituto Nazionale di Astrofisica - Osservatorio Astronomico di Roma, Via Frascati 33, I-00040, Monteporzio Catone, Italy;
marco@mporzio.astro.it}

\altaffiltext{3}{School of Mathematical Sciences, P.O. Box, 28M, Monash University, Victoria 3800, Australia}

\altaffiltext{4}{Centre for Astrophysics and Supercomputing, Swinburne University of Technology, Mail Number 31,
P.O. Box 218, Hawthorn, Victoria 3122, Australia}

\begin{abstract} 

We present a new and homogeneous set of explosive yields for masses 
$\rm 13,~15,~20,~25,~30~and~35~M_\odot$ and metallicities $Z=0$, 
$10^{- 6}$, $10^{- 4}$, $10^{- 3}$, $6\cdot 10^{- 3}$, $2\cdot 10^{-
2}$. A wide network extending up to Mo has been used in all the 
computation. We show that at low metallicities ($\rm Z \le 10^{- 4}$) 
the final yields do not depend significantly on the initial chemical 
composition of the models so that a scaled solar distribution may be 
safely assumed at all metallicities. Moreover, no elements above Zn 
are produced by any mass in the grid up to a metallicity $\sim 10^{-
3}$. These yields are available for any choice of the mass cut upon 
request.

\end{abstract}


%*********************************************  INTRODUCTION   **************************************************

\section{Introduction}

A proper understanding of the chemical evolution of our galaxy and of 
the universe in general requires a good knowledge of the chemical 
composition of the matter ejected by stars of different masses and 
initial composition. Massive stars certainly play a pivotal role in 
the chemical enrichment of the interstellar medium because they are 
very probably responsible for the production of at least most of the 
intermediate mass elements (O through Ca). In spite of their central 
role in the general comprehension of the chemical evolution of the 
matter, only one extended set of models has been published so far: the 
one computed and discussed by Woosley \& Weaver (1995), hereinafter 
WW95, and Timmes, Woosley \& Weaver (1995), hereinafter TWW95. Their 
yields are based on presupernova models computed by assuming, among 
the others, no mass loss, no rotation, a moderate amount of 
overshooting and semiconvection, a value of the $\rm ^{\rm 
12}C(\alpha,\gamma)^{\rm 16}O$ calibrated on preexplosive yields and a 
network extending up to Ge. The explosions were computed in spherical 
symmetry and the yields eventually obtained by imposing the ejecta to 
have a specific final kinetic energy (their cases A, B and C). The 
initial chemical composition of the models at intermediate 
metallicities was obtained by means of a galactic chemical evolution 
model (described by TWW95).

Unfortunately, the present simulations of both the presupernova 
evolution and the explosion are still far from being robustly 
established. Qualitatively (and partly quantitatively) we know how and 
where the various nuclei are synthesized (see, e.g., WW95, Arnett 
1996, Thielemann, Nomoto \& Hashimoto 1996, Limongi, Straniero \& 
Chieffi 2000), but still large uncertainties connected to both the 
hydrostatic evolution and the explosion of massive stars prevent a 
rigorous computation of the yields. Such uncertainties are mainly 
related to the efficiency of the convection (see, e.g., Chiosi \& 
Maeder 1986, Woosley \& Weaver 1988, Bazan \& Arnett 1994), the 
determination of the cross section of a few nuclear processes (first 
of all the $\rm ^{12}C(\alpha,\gamma)^{16}O$ - see, e.g., Weaver \& 
Woosley 1993 and Imbriani et al. 2001), the time delay between the 
collapse of the core and the rejuvenation of the shock wave and the 
precise location of the mass cut (which is the mass coordinate that 
separates the part of the star that collapses in the remnant from the 
one that is ejected outward), even in spherical symmetry. To further 
complicate the situation, also rotation, mass loss, magnetic field and 
asymmetric explosions may also produce large variations in the final 
yields (see, e.g., Heger, Langer \& Woosley 2000 and Maeda \& Nomoto 
2003).

Some years ago we started a long term project devoted to the study of 
the evolution of massive stars and their associated explosive yields 
(Chieffi, Limongi \& Straniero 1998, Limongi, Straniero \& Chieffi 
2000, Limongi \& Chieffi 2002, Chieffi \& Limongi 2002a, Limongi \& 
Chieffi 2003 - LC03). Since the beginning we made a strong effort to 
avoid the use of the various kinds of statistical equilibrium usually 
adopted to determine the chemical evolution of the matter at 
temperatures larger than, roughly, 3 billions degrees. Moreover we 
made an effort to fully couple the integration of the physical 
equations to the ones describing the evolution of the nuclear species 
in order to increase the numerical accuracy. Over the years we 
increased progressively the nuclear network that now extends up to 
Molybdenum. However, similarly to WW95, also our models are still 
computed by neglecting both mass loss and rotation. In our latest 
paper (LC03) of the series we presented our most updated version of 
the hydrostatic code (FRANEC) together to our new hydrodynamic code 
needed to follow the propagation of the blast wave. We also showed 
that the yields produced by a given stellar mass depend mainly on the 
location of the mass cut rather than from the explosion energy. This 
means that, as a first approximation, the yields corresponding to the 
ejection of different amounts of $\rm ^{\rm 56}Ni$ may be obtained by 
assuming an explosion strong enough to eject the full mantle and 
imposing by hand the mass cut at the desired $\rm ^{\rm 56}Ni$ 
abundance. Such a finding means that one can easily explore different 
choices for the mass cut without the necessity of recomputing many 
times the explosion of the models.

By making use of the latest versions of the two codes (hydrostatic and 
hydrodynamic) described in LC03, here we present a wide database of 
yields. In particular, we present the explosive yields produced by a 
grid of six masses ($\rm 13$, $\rm 15$, $\rm 20$, $\rm 25$, $\rm 30$ 
and $\rm 35~M_\odot$) and six metallicities ($\rm Z=0$, $10^{- 6}$, 
$10^{- 4}$, $10^{- 3}$, $6\cdot 10^{-3}$, $2\cdot 10^{-2}$).

The paper is organized as follows. The evolutionary code and the input 
physics adopted to compute the grid are briefly summarized in section 
2. Section 3 is devoted to the discussion of the initial chemical 
composition used to compute the models in the intermediate metallicity 
range between the primordial and the solar one. A final discussion and 
conclusions follow.

\section{The hydrostatic and the hydrodynamic codes}

The presupernova evolutions have been computed by means of the latest 
version of the FRANEC code which has been described in LC03 (and 
references therein). Let us just recall here that the nuclear network 
extends up to Molybdenum and includes 40 isotopes (from neutrons to 
$\rm  ^{30}Si$) in hydrogen burning, 149 isotopes (from neutrons to 
$\rm ^{98}Mo$) in helium burning and 267 isotopes (from neutrons to 
$\rm ^{98}Mo$) in all the more advanced burning phases. In total 282 
isotopes (see Table 1 in LC03) and about 3000 reaction rates were 
explicitly included in the various nuclear burning stages. The nuclear 
network is fully coupled to the equations describing the physical 
structure of the star so that both the physical and chemical evolution 
due to the nuclear reactions are solved simultaneously. No nuclear 
statistical equilibrium (NSE) approximation has been adopted at high 
temperatures.
 
The explosive nucleosynthesis associated to the explosion of each 
massive star model is computed with the same procedure described in 
LC03. The propagation of the shock front through the mantle of the 
star is followed by solving the hydrodynamical equations in spherical 
symmetry and in lagrangean form, following the prescription of 
Richtmeyer \& Morton (1967) and Mezzacappa \& Bruenn (1993). The 
chemical evolution of the matter is computed by coupling the same 
nuclear network adopted in the hydrostatic calculations (Table 1 of 
LC03) to the hydrodynamical equations. The explosion is started by 
imparting an initial velocity $v_{0}$ to a mass coordinate of $\rm 
\simeq 1~M_\odot$ of the presupernova model, i.e. well inside the iron 
core, and by imposing the inner edge of the exploding mantle to move 
on a ballistic trajectory under the gravitational field of the compact 
remnant. $v_{0}$ is properly tuned in order to eject all the mass 
above the Fe core. By taking advantage of the fact that the final 
yields mainly depend on the mass cut location (see LC03), yields 
corresponding to different amounts of $\rm ^{\rm 56}Ni$ ejected are 
then easily obtained by fixing the mass cut by hand a posteriori.

\section{The initial composition of the stellar models}

We computed the presupernova evolution of the six massive star models, 
for various metallicities ranging from zero to solar. The zero 
metallicity models were computed by assuming an initial primordial 
composition ($Z$=0, $Y$=0.23), while the solar metallicity ones were 
computed starting with a scaled solar heavy elements distribution, as 
derived from Anders \& Grevesse (1989), and an initial helium mass 
fraction $Y$=0.285. The initial chemical composition adopted between 
these two extreme metallicities requires some comments. In general, 
the initial composition of a star of a given metallicity is the result 
of the enrichment of the interstellar medium provided by the previous 
stellar generations; hence, its determination would involve a Galactic 
chemical evolution (GCE) model and therefore it would depend on the 
IMF, SFR, Infall, Chemical Yields, etc. Such autoconsistent procedure 
has been adopted by WW95 and TWW95 to determine the initial chemical 
composition of the models of intermediate metallicities ($\rm 
0<Z<Z_\odot$). Unfortunately models computed in this way are obviously 
strictly linked to the GCE model they belong to and should not be used 
in any other GCE simulation. Hence, the computation of the stellar 
models, and their associated explosive yields, should be redone for 
any GCE simulation, and such a procedure would obviously require an 
enormous amount of computer time. It is therefore crucial to 
understand if it is possible (or not) to compute a grid of explosive 
yields of {\it general} purpose and, obviously, which is the initial 
chemical composition that should be used: in the following we will 
address such a problem.

In order to study how (and if) the specific abundances of the various 
nuclei affect the final yields we performed two test evolutions of a 
$\rm 25~M_\odot$ model having an initial global metallicity $Z=10^{-
4}$. We chose this metallicity because the largest deviations from a 
scaled solar distribution occur at low metallicities. In the first 
test we started from a scaled solar distribution and set to zero the 
abundances of all the nuclei but $\rm ^{12}C,~^{14}N,~^{16}O$ and $\rm 
^{56}Fe$. The upper panel in Figure 1 shows the logarithmic ratio 
between the yields obtained in the test case and the standard ones 
(i.e. those obtained with a scaled solar composition). It is quite 
evident that the two sets are in very good agreement. Co is the only 
element that shows a difference by a factor of 2. This test clearly 
demonstrates that the initial abundances of the elements initially set 
to zero do not influence significantly the final explosive yields. 
Hence, for sake of simplicity, we can adopt a scaled solar 
distribution for all of them.

The next step is to understand how (and if) the initial abundances of 
the CNO nuclei affect the final yields. Hence, we performed a second 
test in which, starting from a scaled solar distribution, we imposed a 
[O/Fe] equal to 0.4 dex and a global metallicity $Z=10^{-4}$. 
Obviously this test automatically includes also a possible variation 
of the initial abundances of C and/or N because the initial relative 
abundances among the CNO nuclei are promptly brought to their 
equilibrium values as soon as the star settles on the Main Sequence. 
This test is particularly interesting also because the global 
abundance of CNO controls the size of the H convective core and it is 
the starting point of the important chain $\rm 
^{14}N(\alpha,\gamma)^{18}F(\beta+)^{18}O(\alpha,\gamma)^{22}Ne(\alpha,
n)^{25}Mg$ that is a very important neutrons source. The lower panel 
in Figure 1 shows the logarithmic ratio between the yields obtained in 
the test and in the standard cases: once again the two sets of yields 
are in good agreement (within a factor of 2) even if a few elements 
show now some differences (largely confined, however, within a factor 
of 4): these elements are N, F, K, Sc, Cu and Zn. N is a typical 
product of the H burning and its final yield directly depends on the 
initial CNO abundance. Hence, it is quite obvious that a scaled solar 
distribution can not provide the same yield provided by an initial CNO 
enhanced composition. However, since probably massive stars are not 
the main contributors to the N production in the Galaxy, the adoption 
of an initial scaled solar abundance for N does not constitute a too 
serious problem. F is probably mostly produced by the neutrino induced 
reactions during the explosion (WW95). Since these processes are not 
presently included in our models, our current yield for F is not much 
reliable anyway. Also the differences obtained for K, Sc, Cu and Zn 
should not be considered as a big problem because, in any case, their 
production totally relies on the location of the mass cut (the mass 
location that divides the part of the star that eventually collapses 
in the remnant from the one that it is expelled outward) that is still 
a very uncertain theoretical prediction. Hence, waiting for yields 
based on more reliable explosive models, we conclude that at present 
the adoption of an initial scaled solar distribution of all the 
elements relative to Fe is a reasonable compromise between generality 
and accuracy. Therefore, we will assume in the following a scaled 
solar distribution (Anders \& Grevesse 1989) for all the metallicities 
higher than zero.

The weak dependence of the {\it elemental} yields on the initial 
chemical composition obtained above is not surprising because the most 
abundant isotope of each even element is of primary origin (explosive 
and/or hydrostatic) while the odd elements are always produced by a 
combination of a primary and a secondary component; as the metallicity 
lowers, the secondary component drops to zero but the primary one 
remains obviously active. It goes without saying, at this point, that 
such findings also justify (a posteriori) the use of the WW95 yields 
in GCE simulations other than the one (TWW95) they come from.

The grid of initial metallicities we eventually chose is: 
($Z,Y$)=(0,0.23), ($\rm 10^{- 6}$,0.23), ($\rm 10^{-4}$,0.23), ($\rm 
10^{-3}$,0.23), ($\rm 6\cdot 10^{-3}$,0.26) and ($\rm 2\cdot 10^{-
2}$,0.285), where $Z$ stands for the global metallicity and $Y$ for 
the initial $\rm ^{4}He$ mass fraction.

\section{Discussion and Conclusions}

The final explosive isotopic yields in solar masses of all the 
computed models are reported in Table 1, available only in electronic 
format, once all the unstable isotopes have decayed into their stable 
isobars. The yields of selected radioactive isotopes at $\rm 10^{7}~s$ 
after the explosion are collected in the same table. For obvious 
reasons we could not present different sets of yields for different 
choices of the mass cut; hence we chose to present just one case, i.e. 
the one in which all masses eject $\rm 0.1~M_\odot$ of $\rm ^{\rm 
56}Ni$. Any other choice is promptly available upon request.

The full set of elemental production factors (PFs) is shown in Figure 
2. Each panel refers to a specific metallicity and each symbol refers 
to a given mass (see Figure caption). Let us remind that, in our case, 
the PF of any given isotope/element is defined as the ratio of each 
isotope's/element's mass fraction in the total ejecta divided by its 
corresponding initial mass fraction, i.e.,  ${\rm 
PF}=X_{ejected}/X_{ini}$. Note that this definition is different from 
the one usually adopted by WW95, where ${\rm 
PF}=X_{ejected}/X_{\odot}$.

Some basic properties of the present yields may be seen by looking at 
Figure 2. First of all the PFs of all the elements from C to Zn 
significantly decrease as the metallicity increases, almost 
independently on the initial mass - the only exceptions being N and F. 
The reason for this is that, regardless of the mass of the star, the 
yields of the elements do not vary by more than an order of magnitude 
within the entire range of metallicities (see Table 1), while the 
$X_{ini}$ obviously scale directly with the initial global metallicity 
and hence they vary by several order of magnitudes. Such a strong 
dependence of the PFs on the metallicity simply means that the larger 
the metallicity, the more difficult is the further chemical 
enrichment.

A second feature is the well known {\em odd-even effect}, i.e. that 
the difference between the PFs of the odd (Na to Sc) and the even 
nuclei (Ne-Ca) decreases as the metallicity increases: at the solar 
metallicity most of the elements show a roughly scaled solar 
distribution (see LC03 for a more detailed discussion of this topic). 
It is worth noting that, with the $\rm ^{12}C(\alpha,\gamma)^{16}O$ 
rate adopted in the present calculations, Ne, Mg, Si, S, Ar and Ca 
preserve a scaled solar distribution at all the metallicities (see 
Imbriani et al. 2001 for a more comprehensive discussion of this 
topic).

A last feature worth mentioning here is that below $Z=10^{-3}$ there 
is a cutoff in the PFs at the level of Zn, i.e., no elements heavier 
than Zn are produced. On the contrary, above $Z=10^{-3}$ such a cutoff 
progressively reduces so that a consistent production of elements 
beyond Zn is obtained. Elements above Sr are not produced in a 
significant amount even at solar metallicity. This means that the 
observed abundances of elements above Zn in very metal poor stars must 
be attribute to stars (or, in general, to processes) outside the range 
presently analyzed.

Since the only other paper presenting a full set of yields is the WW95 
one, we show in Figure 3 the comparison between the WW95 and the 
present yields, for two masses and three metallicities. Only elements 
up to Ge are shown because the nuclear network adopted by WW95 does 
not extend above this element. For this comparison we chose, for each 
stellar model, the mass cut that provides the ejection of the same 
amount of $\rm ^{\rm 56}Ni$ as in the corresponding WW95 model. Note 
that, since the grid of metallicities computed by WW95 does not 
coincide exactly with the ones presented here, the comparison shown in 
Figure 3 refers to models having a slightly different initial 
metallicity. We selected the $\rm 20$ and the $\rm 25~M_\odot$ because 
are the ones that dominate the yields of a stellar generation having a 
Salpeter like IMF (see, e.g., LC03). The lower right panel in Figure 3 
shows that there is a very good agreement between ours and the WW95 
yields for the $\rm 25~M_\odot$ of solar metallicity. On the contrary, 
all other panels disclose significant (and not systematic!) 
differences between the two sets of yields. In particular there are a 
few things worth noting: a) both sets of models produce O and C 
in similar amounts (within a factor of two) while the N yields tend to 
be similar only for $\rm Z\ge10^{-3}$, b) the light elements Ne, Na 
and Mg tend to be significantly more produced in our models than in 
the WW95 ones while Al is produced in quite similar amounts, c) we 
tend to systematically underproduce the products of the explosive 
oxygen burning and incomplete Si burning, i.e. Si, S, Ar and Ca by 
roughly a factor of two with respect to WW95 (even if the relative 
scaling among these elements is remarkably similar), d) also the odd 
elements P, Cl and K, tend to be quite largely underproduced in our 
models with respect to WW95 and e) the Iron peak nuclei show a quite 
contradictory behavior because, while Ti is always in good agreement, 
Co and Ni are generally overproduced and Sc often underproduced 
relative to WW95.

A proper understanding of the sources of such differences, though of 
overwhelming interest, is extremely difficult because the chemical 
yields are, in general, the result of a complex interplay among the 
various hydrostatic evolutionary phases plus the subsequent passage of 
the shock wave (Chieffi, Limongi \& Straniero 2000). For example, 
elements like N and Mg are not significantly affected by the passage 
of the shock wave and hence their final differences will mainly 
reflect a different presupernova evolution (but note that, e.g., O, 
that it is also a product of the hydrostatic burnings, is produced in 
very similar amount). Other elements are produced, viceversa, only by 
the explosive burnings and therefore one could think that playing with 
the mass cut could significantly improve the comparison. This is not 
the case. First of all let us note that the mass cut must be located 
within the region undergoing complete explosive Si burning because 
appreciable amounts of Sc, Co and Ni must be ejected. Hence the 
abundances of the elements produced by the explosive oxygen burning 
and/or incomplete explosive Si burning (Si, S, Ar, K, Ca, V, Cr and 
Mn) would not be modified by a changing of the mass cut. But also the 
comparison of the elements mainly produced by the complete explosive 
Si burning would not be improved by a changing of the mass cut because 
a better fit to any of the elements like Sc, Ti, Co and Ni would 
worsen the fit to the others. 

A deeper comparison between these two sets is virtually impossible 
because either the two sets of models have been computed by adopting 
different choices for both the treatment of the convective layers and 
the rate of the $\rm ^{12}C(\alpha,\gamma)^{16}O$ nuclear process, and 
also because the models on which the WW95 yields are based have never 
been published. The only possible comparisons between our presupernova 
models and the ones that are at the base of the WW95 yields have been 
presented in Limongi, Straniero \& Chieffi (2000) and hence we refer 
the reader to that paper for such a comparison.

The differences between the WW95 and our yields are large enough that 
they should produce visible differences in GCE simulations and hence 
we strongly suggest the use of both sets of yields in the GCE 
modeling so to understand how alternative sets of yields influence 
our current understanding of the chemical evolution of the universe.

In conclusion, we provide in this paper a brand new set of yields in a 
wide range in both mass and initial metallicity. All the yields are 
freely available to the community for any choice of the mass cut (upon 
request). We have shown for the first time that the initial chemical 
composition does not affect significantly the final yields up to at 
least a metallicity of the order of $Z=10^{-4}$. We have also shown 
that a metallicity larger than $Z=10^{-3}$ is necessary to begin to 
produce elements beyond Zn up the neutron magic number $\rm N=50$. The 
present yields are quite different from the WW95 ones and the observed 
differences cannot be simply explained in terms of one or few causes 
but are certainly due to the complex interplay among various aspects 
of both the hydrostatic evolution and the explosion itself that are 
very difficult to disentangle at the moment.

\acknowledgements

A.C. warmly thanks John Lattanzio and Brad Gibson for their kind 
hospitality in Melbourne and for having provided the computer 
facilities (the Australian Partnership for Advanced Computing National 
Facility and the Swinburne Centre for Astrophysics and Supercomputing 
in Melbourne) necessary to perform such a huge amount of computations.

\begin{figure} % fig.1
\plotone{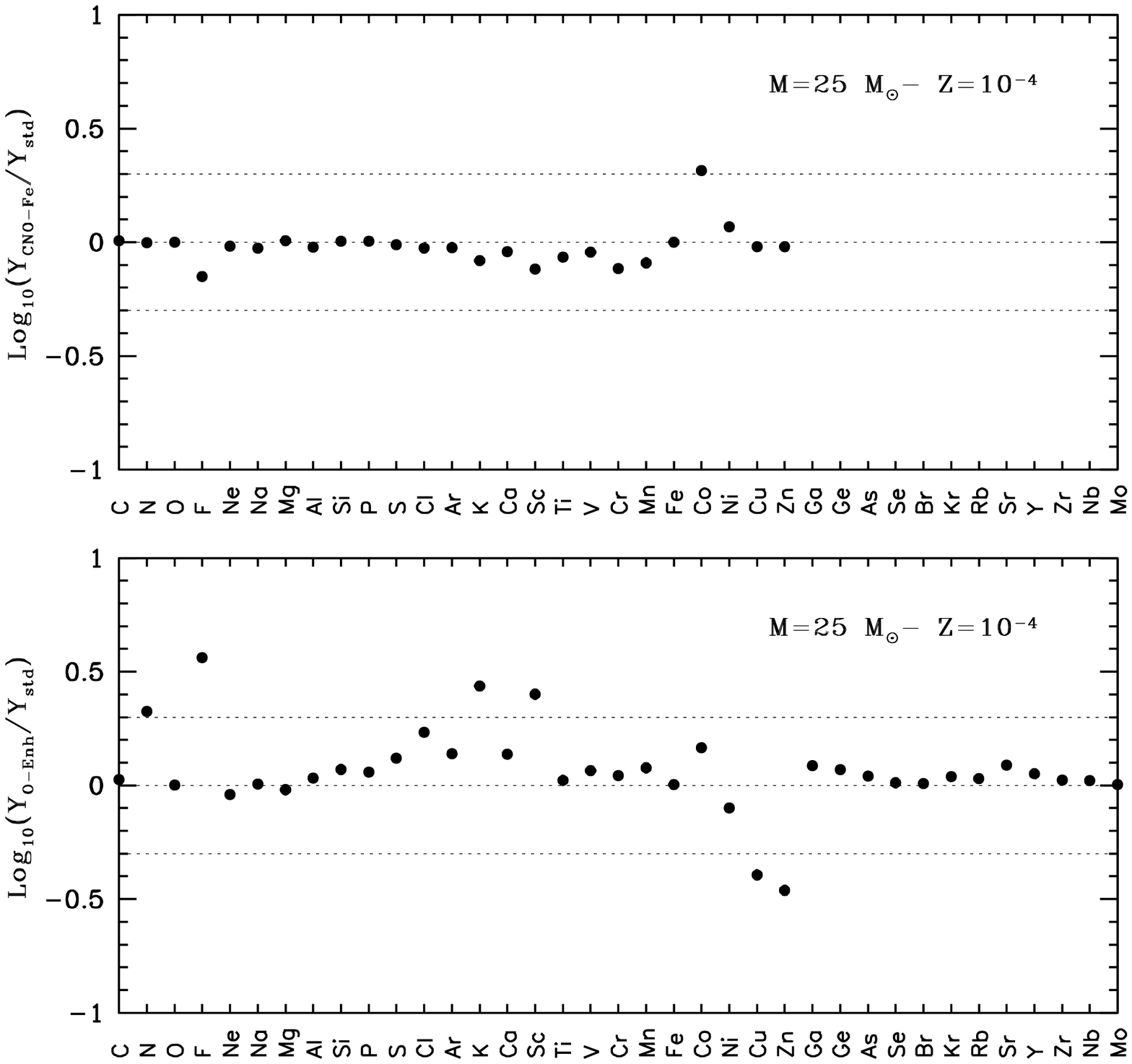}

\caption{{\em Upper panel}: Logarithmic ratio between the explosive yields produced 
by a $\rm 25~M_\odot$ of global metallicity $Z=10^{-4}$ in which the initial 
abundances of all the nuclei are set to zero with the exclusion of those of $\rm 
^{12}C$, $\rm ^{14}N$, $\rm ^{16}O$ and $\rm ^{56}Fe$ and the ones produced by a 
standard $\rm 25~M_\odot$ having an initial scaled solar metallicity $Z=10^{-4}$. 
{\em Lower panel}: Logarithmic ratio between the explosive yields produced by a 
$\rm 25~M_\odot$ of global metallicity $Z=10^{-4}$ and [O/Fe]=0.4 and those 
produced by the standard reference model.
}

\end{figure}

\begin{figure} % fig.2
\plotone{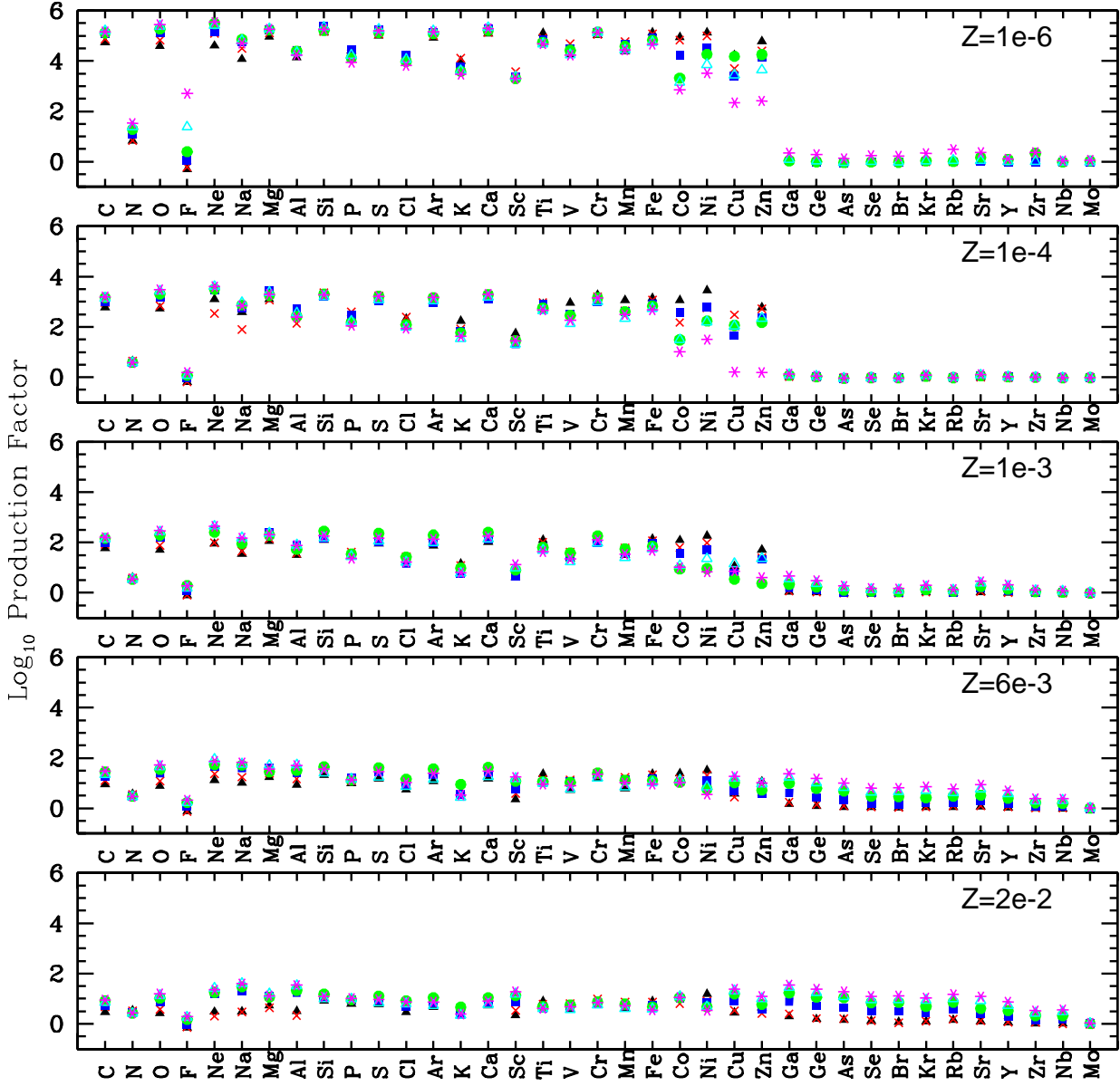}
\caption{Production factors of all the elements from C to Mo. The symbols refer to the 6 
masses: $\rm 13~M_\odot$ ({\em black filled triangles}), $\rm 15~M_\odot$ ({\em red crosses}),
$\rm 20~M_\odot$ ({\em blue filled squares}), $\rm 25~M_\odot$ ({\em 
green filled circles}), $\rm 30~M_\odot$ ({\em cyan open triangles}), $\rm 35~M_\odot$ 
({\em magenta asterisks}). Each panel refers to the metallicity reported in the
upper right corner.
%\label{}
}
\end{figure}

\begin{figure} % fig.3
\plotone{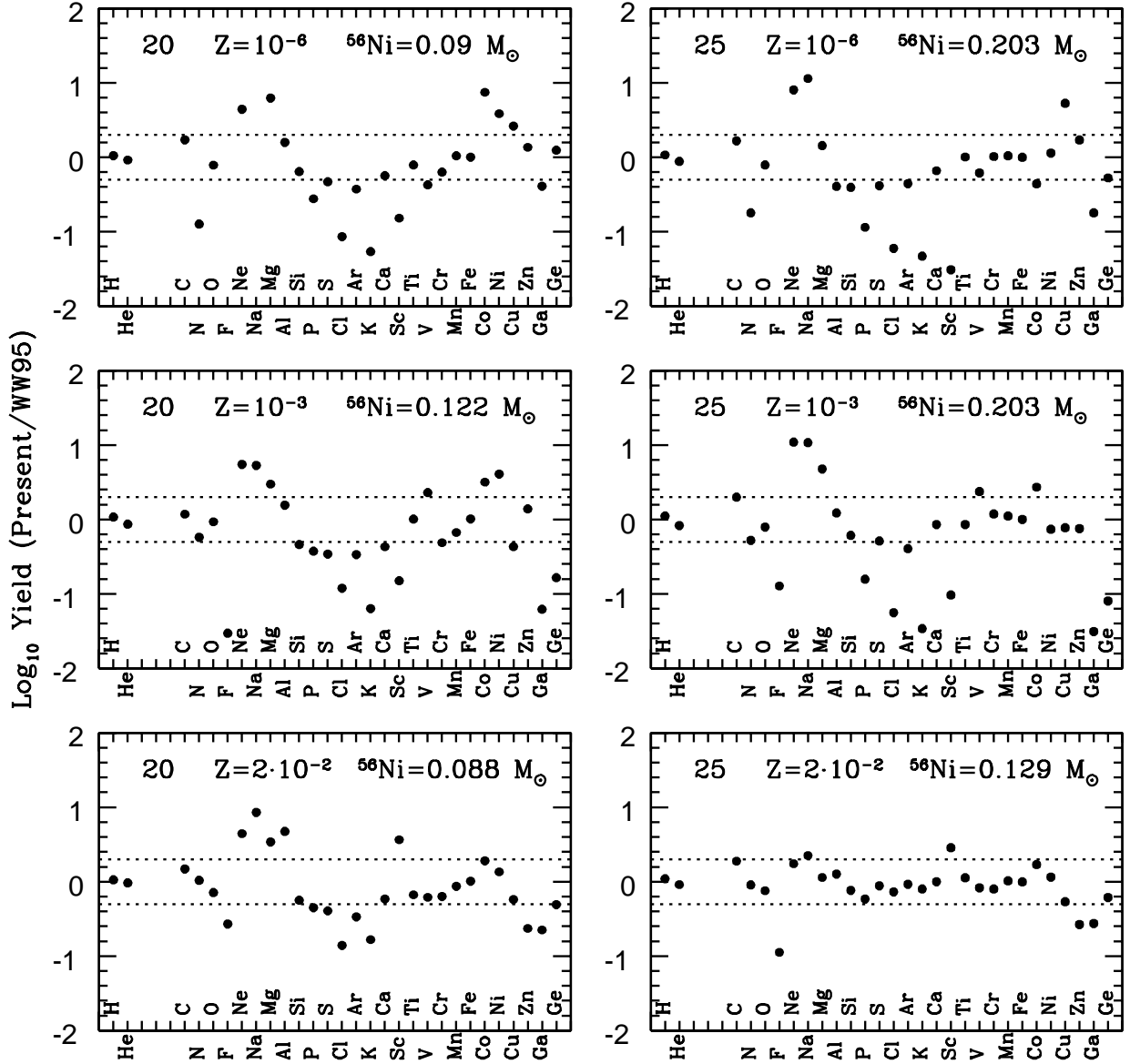}
\caption{Comparison between the elemental yields
provided by WW95 and the present ones for two masses, 20 and $\rm 25~M_\odot$, and
for three selected metallicities, $Z=10^{-6}$ ({\rm upper panels}),
$Z=10^{-3}$ ({\rm middle panels}) and $Z=2\cdot 10^{-2}$ ({\rm lower panels}).
}
\end{figure}

\clearpage

\begin{deluxetable}{lrrrrrr}\label{tabnet}
\tabletypesize{\small}
%\tablenum{3}
\tablewidth{0pt}
\tablecaption{Explosive yields}
\tablehead{
Mass               &13          & 15        &  20        &   25       &    30      &    35   
}
\startdata
               \multicolumn{7}{c}{Metallicity Z=0}                                             \\
\cline{1-7}											\\
$\rm M_{ejected}$  &11.79       & 13.53     &  18.33     &   23.40    &    27.93   &    32.97   \\ 
$\rm M_{remnant}$  &1.21        & 1.47      &  1.67      &   1.60     &    2.07    &    2.03    \\
$\rm ^{56}Ni$      &0.0913      & 0.1000    &  0.1000    &   0.1000   &    0.1000  &    0.1000  \\  
$\rm ^{1 }H $	   & 6.86(+00) & 7.77(+00)  & 9.74(+00)  & 1.16(+01)  & 1.32(+01)  & 1.47(+01)  \\
$\rm ^{2 }H $      & 6.55(-17) & 5.95(-17)  & 6.34(-17)  & 1.77(-16)  & 9.76(-16)  & 3.05(-16)  \\
$\rm ^{3 }He$      & 3.07(-05) & 3.01(-05)  & 2.90(-05)  & 2.77(-05)  & 2.72(-05)  & 2.71(-05)  \\
\enddata            
\end{deluxetable}

\end{document}